\documentclass[pra,twocolumn,amsmath,amssymb,showpacs,superscriptaddress]{revtex4-1}
\pdfoutput=1 
\usepackage{graphicx}
\usepackage{dcolumn}
\usepackage[mathlines]{lineno}
\usepackage{hyperref}
\usepackage{bm}
\usepackage{epstopdf}
\usepackage{epsfig}
\usepackage{bbold}
\usepackage{soul}

\newcommand{\bra}[1]{\ensuremath{\left\langle#1\right|}}
\newcommand{\ket}[1]{\ensuremath{\left|#1\right\rangle}}
\newcommand{\braket}[2]{\ensuremath{\left\langle #1 | #2 \right\rangle}}
\usepackage{color}

\begin{document}

\title{A hybrid quantum eraser scheme for characterization of free-space and fiber communication channels}

\author{Isaac~Nape}
\affiliation{School of Physics, University of the Witwatersrand, Private Bag 3, Wits 2050, South Africa}
\author{Charlotte Kyeremah}
\affiliation{Laser and Fibre Optics Centre, Department of Physics, University of Cape Coast, Cape Coast, Ghana}
\author{Adam~Vall\'{e}s}
\email[Corresponding author: ]{adam.vallesmari@wits.ac.za}
\affiliation{School of Physics, University of the Witwatersrand, Private Bag 3, Wits 2050, South Africa}
\author{Carmelo~Rosales-Guzm\'an}
\affiliation{School of Physics, University of the Witwatersrand, Private Bag 3, Wits 2050, South Africa}
\author{Paul K. Buah-Bassuah}
\affiliation{Laser and Fibre Optics Centre, Department of Physics, University of Cape Coast, Cape Coast, Ghana}
\author{Andrew~Forbes}
\affiliation{School of Physics, University of the Witwatersrand, Private Bag 3, Wits 2050, South Africa}
\date{\today}

\begin{abstract}
\noindent We demonstrate a simple projective measurement based on the quantum eraser concept that can be used to characterize the disturbances of any communication channel. Quantum erasers are commonly implemented as spatially separated path interferometric schemes. Here we exploit the advantages of redefining the \emph{which-path} information in terms of spatial modes, replacing physical paths with abstract paths of orbital angular momentum (OAM). Remarkably, vector modes (natural modes of free-space and fiber) have a non-separable feature of spin-orbit coupled states, equivalent to the description of two independently marked paths. We explore the effects of fiber perturbations by probing a step-index optical fiber channel with a vector mode, relevant to high-order spatial mode encoding of information for ultra-fast fiber communications.
\end{abstract}

\maketitle


\section{Introduction}

The concept of \emph{which-way} information has profound implications in the study of coherence of light, giving a different scope to the historical  wave-particle duality. The most revered demonstration of the wave-like nature of photons was performed by Young in 1804, in his famous double-slit experiment \cite{young1804}, followed by modern variations \cite{rauch1974test, grangier1986experimental, zeilinger1988single,taylor1909interference,gerlich2011quantum}. The double-slit experiment is a two-path interferometer in which a light source blocked by a screen with two slits is split into two new sources traveling along different paths. Upon propagation, these two new sources interfere with each other to produce an interference pattern of spatial fringes. Remarkably, interferometric phenomena are not restricted to two-path interferometers; it is also possible to observe interference of beams traveling along the same path, but using different degrees of freedom (DoF). 

The combination of two DoF, orbital angular momentum (OAM) and polarisation in a non-separable fashion, known as vector beams, allows to perform novel versions of the two-slit experiments. Here, the physical paths are replaced by two components of one degree of freedom, e.g., two values of OAM. Vector beams have gained significant amount of interest in a great variety of research fields at both the classical and the quantum levels. In particular, in the field of optical communication and quantum information, their high dimensional encoding capabilities have raised attention \cite{nagali2010experimental,milione20154,sit2016high, nape2016high} due to their potential applications in free-space and optical fibers \cite{gregg2015q, ramachandran2009generation, ndagano2015fiber}. In quantum optics, photons entangled in OAM and polarisation have been demonstrated to violate a Bell-like inequality \cite{karimi2010}, being able also to tune its entanglement or photon indistinguishability \cite{valles2014}, similarly to the analogous version of a quantum eraser scheme using OAM and polarisation \cite{nape2017erasing}. Other DoF can also be found to demonstrate this particular type of correlations, e.g., in the case of generating entanglement between momentum and polarisation in a single photon \cite{gadway2008bell}, or even using intense beams \citep{borges2010,kagalwala2013}.

The modern view of wave-particle duality has opened new research avenues, for example in the development of novel measurement schemes, as the ones based on quantum non-demolition \cite{braginsky1980}. The traditional quantum eraser experiment \cite{scully1982quantum,scully1991quantum,walborn2002double, neves2009control,neves2009hybrid, kwiat1992observation,herzog1995complementarity, kim2000delayed, ma2013quantum}, and its delayed choice versions \cite{chen2014revisiting,ma2016delayed, ma2013quantum, jacques2008delayed} are related to the complementarity principle formulated by Bohr in 1928 \cite{bohr1928quantum}, which states that photons can behave indistinctly as particles or waves but cannot be observed as both simultaneously. 

Importantly, the double-slit experiment and its modern variations allows to link the \emph{which-way} information provided by the whole system with the interference pattern produced at the detection plane \cite{zou1991}. Thus, the visibility of the interferometric pattern can be directly related to the properties of the system. For example, the decrease of quality in the interferometric measurement can be associated to the perturbations introduced by the system. This approach provides a useful tool for applications in optical communication in both free-space and optical fibers \cite{gregg2015q, ramachandran2009generation, ndagano2015fiber}, a hot topic nowadays due to the realization of a pending bandwidth ``capacity crunch".

Here we report on the comparison of a quantum eraser in free-space and a step-index fiber using the OAM and polarization DoF provided by vector modes. This work establishes the basis for a simple detection technique to quantify perturbations introduced by the environment, particularly in the communication channel, between the source and the detection section. In our particular case, a fiber optic link is formed by many different channels, introducing some kind of perturbation depending on the DoF that carries the encoded information. That is to say, the results presented give a fast probing method to characterize the different channel's perturbation, and would be useful for determining in an easy measurement what would be feasible for a quantum communication channel down fiber. 

\section{Concept}

\subsection{Revisiting the traditional \emph{which-way} quantum eraser}
Previous demonstrations of the quantum eraser experiment were performed with the aid of path interferometers. We revisit a variation of the experiment based on Thomas Young's double-slit interferometer. Single photons traversing the two slits form  interference fringes due to each photon's paths interfering (wave-like behaviour). However, the lack of interference fringes is associated with path distinguishability (particle-like behaviour). In the quantum eraser experiment, the interference fringes and \emph{which-way} information cannot be observed simultaneously. Quantitatively, this is associated with the complementarity inequality \cite{wootters1979complementarity, greenberger1988simultaneous, jaeger1995two,englert1996fringe},
\begin{equation}
	V^2+D^2 \leq 1,\label{eq:complimentary}
\end{equation}
where $D$ is the amount of path information in the system while $V$ is the visibility of interference fringes. Thus, gaining knowledge of path information ($D \neq 0$), reduces the visibility of the fringes ($V < 1$). In the quantum eraser experiment, the path information can be obtained (minimal V) and subsequently erased (maximal V). To illustrate this effectively with the aid of Eq. (\ref{eq:complimentary}), consider a double-slit marked with orthogonal polarizers. The quantum state of the system, is given by
 \begin{equation}
 \ket{\Phi}=\frac{1}{\sqrt{2}}\left(\ket{H}\ket{s_{1}}+\ket{V}\ket{s_{2}}\right),\label{eq:markedSlits}
 \end{equation}
with $\ket{s_1}$ and $\ket{s_2}$ the states upon traversing the independent paths $s_1$ and $s_2$, respectively, and $\ket{H}$ and $\ket{V}$ represent the horizontal and vertical polarization states that mark the two paths. Note that without the markers and taking into account perfect conditions, the two paths are allowed to interfere, which leads to the trivial case of interference fringes appearing at the detection plane, with $D=0$ (minimal path information) and $V=1$ (maximal fringe visibility), due to path indistinguishably (wave-like behavior). On the contrary when the slits are marked, the probability distribution of the photons is $|\braket{\Phi}{\Phi}|^{2} = \sum_{i} |\braket{\psi_i}{\psi_{i}}|^2/2$, which signals the presence of path information in the system when projecting the polarization of the system onto the \ket{H} or \ket{V} states. Thus, $D=1$ (maximal path information) and $V=0$ (minimal fringe visibility), meaning that there is a full knowledge of the \emph{which-path} information (particle-like behavior).

However, the interference fringes can be recovered with a complimentary projection of the polarisation, in the diagonal basis ($\ket{H} \pm \ket{V}$), which acts to remove the path information and hence erase it from the system. Again, $D=0$ (minimal path information) and $V=1$ (maximal fringe visibility), showing a mutually exclusivity between the two cases. Intriguingly,  partial visibility and partial distinguishability are permitted, where the result cannot be explained exclusively by a wave-like or particle-like interaction although the inequality in Eq.~(\ref{eq:complimentary}) is maintained \cite{englert1996fringe}.

\subsection{Redefining the quantum eraser with spatial modes}

Equation~(\ref{eq:markedSlits}) represents a general state of a non-separable or entangled path and polarisation DoF of a single photon, a trait of non-separable DoF of a photon \cite{gadway2008bell}. Similarly, vector modes are a  class of spatial modes with non-separable polarization and OAM DoF with the following general form
\begin{equation}
\ket{\psi} = \frac{1}{\sqrt{2}} \big(\ket{R}\ket{\ell} + e^{i\zeta }\ket{L}\ket{-\ell}\big). \label{eq:VectorMode}
\end{equation}

\begin{figure*}[ht!]
	\centering
	\includegraphics[width=1\linewidth]{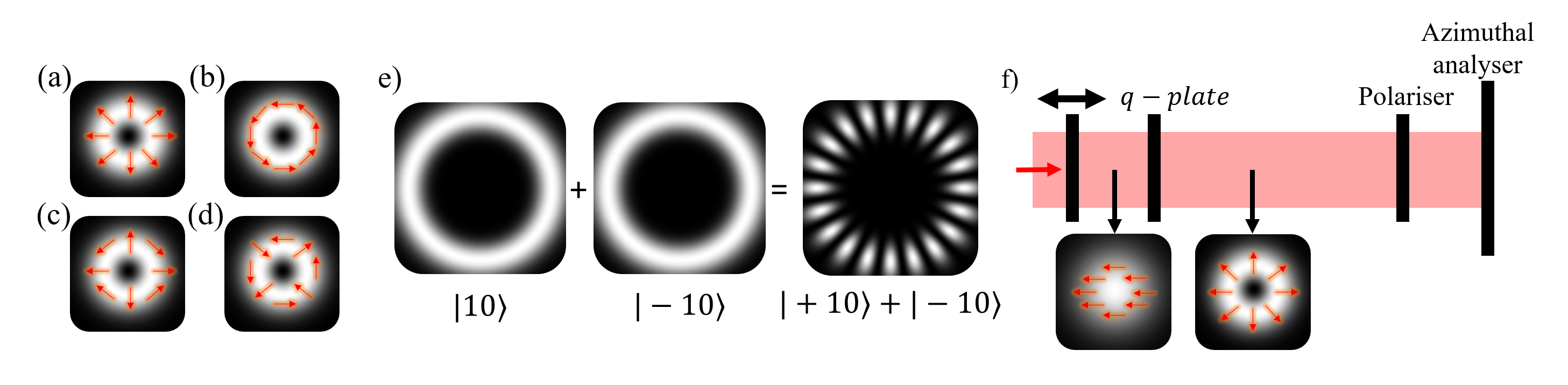}
	\caption{Intensity profile of vector modes described by Eq.~(\ref{eq:VectorMode}), belonging to the $\ell=1$ subspace. (a) is the $\text{TM}_{01}$ mode ($\zeta=0$, $\ell=1$) with field vectors pointing radially, (b) is the $\text{TE}_{01}$ mode ($\zeta=\pi$, $\ell=1$) with azimuthal field lines, (c) is the hybrid electric even  $\text{HE}_{21}^\text{even}$ ($\zeta=0, \ell=-1$) and (d) hybrid odd $\text{HE}_{21}^\text{odd}$ ($\zeta=\pi, \ell=-1$) modes. (e) is an illustration of the appearance of azimuthal fringes when two OAM modes with a topological charge $\ell=\pm10$ are in a superposition state. The fringes appear in the azimuthal direction with a frequency of $2|\ell|$. (f) is an illustration of the OAM quantum eraser where the non-separable spin-OAM coupling would be performed with a $q$-plate. The OAM would be distinguished and erased by a polarisation analyser, where the spatial fringes are analysed by an azimuthal scanner of choice.}\label{fig:VM}
\end{figure*}

Here, $e^{i\zeta }$ is a relative phase, the states $\ket{\pm \ell}$ are the OAM eigenstates with $\ell$  representing the topological charge of the spatial field, characterized by a helical phase of $e^{i\ell\phi}$, and $\ket{R}$ and $\ket{L}$ the right and left circular polarization states, respectively. In Eq.~(\ref{eq:VectorMode}) the OAM  eigenstate of the photon are marked with orthogonal circular polarization states. Through polarization control, the OAM information can be determined and erased. For example, by projecting the photon onto the polarization state $\ket{R}$ or $\ket{L}$, the photon collapses onto the state $\ket{\ell}$ or $\ket{-\ell}$, respectively ($D=1, V=0$), where the spatial fields are azimuthal donut-like rings with opposite helicities. An example is illustrated in Fig.~\ref{fig:VM}(e) for $\ell=\pm10$. Analogously, the OAM eigenstates act as abstract paths in contrast to the double-slit experiment. The OAM modes can be interfered with a complimentary projection of the polarization, i.e, $\ket{R}\pm\ket{L}$, thus collapsing the spatial mode onto a superposition state, $\ket{\ell}\pm\ket{-\ell}$, where the interference fringes appear in the azimuthal direction with a frequency proportional to $2| \ell|$ (see Fig.~\ref{fig:VM}). Hence this erases the OAM information of the photon ($D=0, V=1$). Accordingly, the non-separability is exploited to demonstrate the quantum eraser with a single photon described by a vector mode. Interestingly, these spatial modes are natural modes of free-space and fiber, the basic media of quantum information and communication.

\subsection{Vector mode propagation in step-index fibers}

\begin{figure}[h!]
	\centering
	\includegraphics[width=1\linewidth]{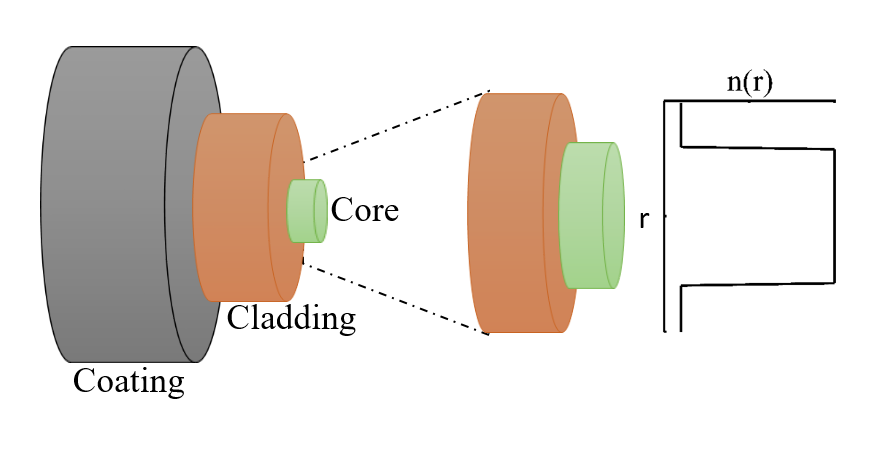}
	\caption{A graphical illustration of the step-index fiber with a characteristic step-like refractive index, where the highest index of refraction is at the core and the lowest is found in the cladding. }\label{fig:fibreSchematics}
\end{figure}

Step-index fibers have cylindrical symmetry and a refractive index with a step-like profile, as can be seen in Fig.~\ref{fig:fibreSchematics}. The full vector wave equation for a step-index fiber is given by
\begin{equation}
\{ \nabla_{t}^{2} + n^2k^2\ \nabla_{t} \} \textbf{u}_t + \nabla \{  \textbf{u}_t \cdot \nabla_t{\text{ln}(n^2)} \}  = \beta^2\textbf{u}_t, \label{eq:Helmholtz}
\end{equation}
where $k=2\pi/\lambda$ is the wave vector, $n$ is the index of refraction which has a radial dependence,  $\textbf{u}_t$ is the transverse component of the electric field while $\beta$ is the propagation constant for each solution. The radial component of the fields are described as follows,
\begin{equation}
 u_{\ell p}(r) =\begin{cases}
	J_{|\ell|}(\frac{\beta_{\ell p}}{a})/J_{|\ell|}(\beta_{\ell p}) &{ r<a},\\[2ex]
 K_{|\ell|}(\frac{\sigma_{\ell p}}{a})/J_{|\ell|}(\sigma_{\ell p})	&{r \geq a},
\end{cases} \label{radialField}
\end{equation}
with $a$ being the fiber core radius and the functions $J_{|\ell|}$ and $K_{|\ell|}$ representing the higher-order Bessel and modified Bessel functions. The $\beta_{\ell p}$ and $\sigma_{\ell p}$ are the respective propagation constants for the Bessel functions in the different regions of the fiber. Equation~(\ref{radialField}) is a consequence of the step-index fibers cylindrical symmetry and refractive index profile.
The first four cylindrically symmetric higher-order vector solutions, which are nearly degenerate, take the form of Eq.~(\ref{eq:VectorMode}). They are known as the transverse electric ($\text{TE}_{01}$), transverse magnetic ($\text{TM}_{01}$), hybrid electric odd ($\text{HE}^{\text{odd}}_{21}$) and hybrid electric even ($\text{HE}^{\text{even}}_{21}$), where the two indices represent the number of half-wave patterns across the width and the height of the waveguide, respectively.

In this paper, we consider the propagation of the $\text{TM}_{01}$ mode which is also known for its radial polarization profile and is defined by Eq.~(\ref{eq:VectorMode}) for $\ell=1$ and $\zeta=0$ (see Fig.~\ref{fig:VM} for intensity profiles).

\section{Implementation}

To generate the spatial modes marked with orthogonal polarisation states we make use of a $q$-plate \cite{marrucci2006optical, marrucci2011spin}, a Pancharatnam-Berry phase element with a locally varying birefringence. A $q$-plate couples the polarization and OAM DoF of light according to the following rules:
\begin{eqnarray}
\ket{R}\ket{\ell}\xrightarrow{q\text{-plate}} \ket{ L}\ket{\ell - 2q}, \label{eq:Qplate1}\\
\ket{L}\ket{\ell}\xrightarrow{q\text{-plate}} \ket{ R }\ket{\ell + 2q},
\label{eq:Qplate2}
\end{eqnarray}
with $\ket{ L }$ and $\ket{ R }$ being the left and right circular polarization states while $q$ represents the charge of the $q$-plate. The spin component of the photon is inverted with an addition in OAM of $\pm2q$. For example, a photon (or an intense beam) with a Gaussian transverse distribution given by $\ket{R}\ket{0}$ is  converted to $\ket{L}\ket{-1}$, if $q=0.5$ according to Eq. \ref{eq:Qplate2}. Notably, the quantum state $\ket{L}\ket{-1}$ corresponds to a scalar mode: a class of spatial modes with separable product states of polarization and OAM DoF. In contrast, vector modes are superpositions of these modes where on the contrary, the OAM and polarization entities are non-separable. These modes are generated by first preparing linearly polarized photons, for example, in the state $\frac{1}{\sqrt{2}}\big(\ket{L}\ket{0} + \ket{R}\ket{0}\big)$, set accordingly with polarization optical elements. The state of the photon upon traversing the $q$-plate is given by,
\begin{equation}
\ket{\psi}=\frac{1}{\sqrt{2}} \big( \ket{R}\ket{\ell} + \ket{L}\ket{-\ell}\big ), \label{eq:hybrid1}
\end{equation}
where $\ell=2q$. Equation~(\ref{eq:hybrid1}) reminds us of the quantum state represented by two paths distinguished by orthogonal polarization markers (see Eq.~(\ref{eq:markedSlits})).
We carry out the  required projections for the quantum eraser on a photon encoded with the state presented in Eq.~(\ref{eq:hybrid1}) through polarization control, followed by a pattern sensitive scanning technique.
To achieve this, firstly we convert the polarization of the photon from the circular to linear basis with a quarter wave-plate (QWP) oriented at $45^\circ$ with respect to the horizontal. Equation (\ref{eq:hybrid1}) now takes the form
\begin{equation}
\ket{\psi}=\frac{1}{\sqrt{2}} \big( \ket{H}\ket{\ell} + e^{i\delta}\ket{V}\ket{-\ell}\big ), \label{eq:hybrid2}
\end{equation}
with a relative phase $e^{i\delta}$ introduced by the QWP. Secondly, a polarization analyzer orientated at an angle $\alpha$ (with respect to the horizontal), will project onto the following target state
\begin{equation}
\ket{\alpha}=\cos(\alpha)\ket{H}+\sin(\alpha)\ket{V}
\label{eq:alpha},
\end{equation}
thus allowing the ``path'' to evolve from marked to unmarked by a judicious choice of $\alpha$.  Next, the visibility of spatial fringes needs to be detected, which may easily be done with scanning detectors (or more expensive camera-based systems). We instead make use of scanning holograms and a fixed detector as our pattern sensitive detector \cite{forbes2016creation}. We create sector states from superpositions of OAM with a relative intermodal phase of $\theta$,
\begin{equation}
\ket{\theta}=\big(\ket{\ell}+e^{i2\theta}\ket{-\ell}\big).
\label{eq:theta}
\end{equation}

The phase structure of $\ket{\theta}$ is azimuthally periodic, and allows a measurement of the path (OAM) interference, analogous to detecting OAM entanglement with Bell-like measurements \cite{leach2009violation, dada2011experimental, fickler2012quantum, mclaren2012entangled}. Thus the fringe pattern (or lack thereof) can be detected by scanning through $\theta$. 

The normalized probability of photon detection given the two projections is 
\begin{align}
	P(\alpha, \theta)&\propto|\bra{\theta}\bra{\alpha}\ket{\psi}|^{2} \nonumber\\
	&=\frac{1}{2}(1+\sin(2\alpha)\cos(2\theta+\delta)).
	\label{eq:prob}
\end{align}

Here $P(\alpha, \theta)$ is equivalent to the experimental photon counts. When the polarizer is orientated at $\alpha=0^\circ$, which corresponds to the $\ket{H}$ polarization state, the probability distribution with respect to $\theta$ is a constant since the OAM abstract path is marked.  Conversely, for $\alpha =\pm45^\circ$ which corresponds to complimentary polarization projections on $\ket{H}\pm\ket{V}$, yields $P(\alpha=\pm45^\circ, \theta)\propto1\pm\cos(2\theta+\delta)$ and hence the sinusoidal dependence is an indication of an interference pattern emerging from a superposition of the OAM abstract paths. Therefore the OAM equivalent \emph{which-path} information has been erased. The fringe visibility is given by
\begin{equation}
V=\frac{P_{max}-P_{min}}{P_{max}+P_{min}},
\end{equation}
where $P_{max}$ and $P_{min}$ are the maximum and minimum photon counts from rotating the azimuthal spatial mode analyzer (SLM). The theoretical visibility of the spatial fringes with respect to the angle of the polarizer ($\alpha$) is  $V=|\sin(2\alpha)|$.

\section{Experimental set-up}

\begin{figure}[t]
	\centering 
	\includegraphics[width=8cm]{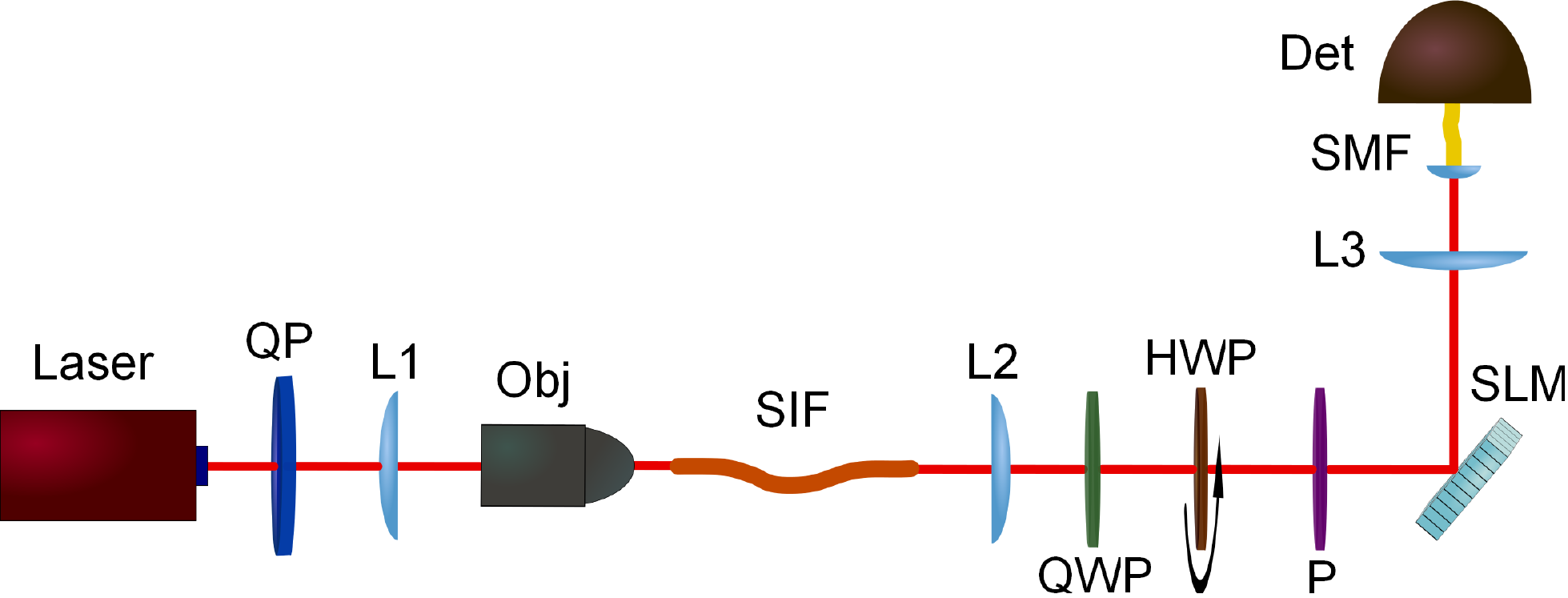}
	\caption{Schematic of the experimental set-up used for the free-space and fiber \emph{which-way} erasure. An  attenuated He-Ne laser source that produces a horizontally polarized Gaussian beam with a nominal wavelength of $633$ nm was used. A $q$-plate (QP) was used to generate the TM$_{01}$ (radial mode) and subsequently imaged into a 30 $\mu$m diameter core step-index fiber (SIF) with lens L1 and a $20\times$ microscope objective (Obj). Lens L2 collect the photons at the output of the fiber. The half wave-plate (HWP) and quarter wave-plate (QWP) are used for polarization control while the spatial light modulator (SLM) carries the spatial scanning. This is imaged into a single-mode fiber with lens L3 and fiber coupler. Note that free-space experiment is carried out in absence of Obj, SIF and L2.}		
	\label{fig:Setup}
\end{figure}

We illustrate the concept of both the generation and detection of our intra-particle quantum eraser with vector modes in free-space and step-index fiber, with the aid of Fig.~\ref{fig:Setup}. In our experiment, we made use of an attenuated laser as the single-photon source. The incident beam was horizontally polarized in order to be able to generate our vector mode, by means of correlating the polarization with the OAM DoF using a $q$-plate {($q$ = 0.5)} \cite{marrucci2006optical, marrucci2011spin}. The resulting state after the $q$-plate is given by Eq.~(\ref{eq:hybrid1}), and is the one coupled into the step-index fiber as seen schematically in Fig.~\ref{fig:fibreSchematics}. In order to give a proper analysis of the hybrid state disturbance introduced by the step-index fiber, we first performed the projective measurement to the free-space transmitted vector mode to have reference curves for the best case scenario.

\begin{figure}[t]
	\centering
	\includegraphics[width=8 cm ]{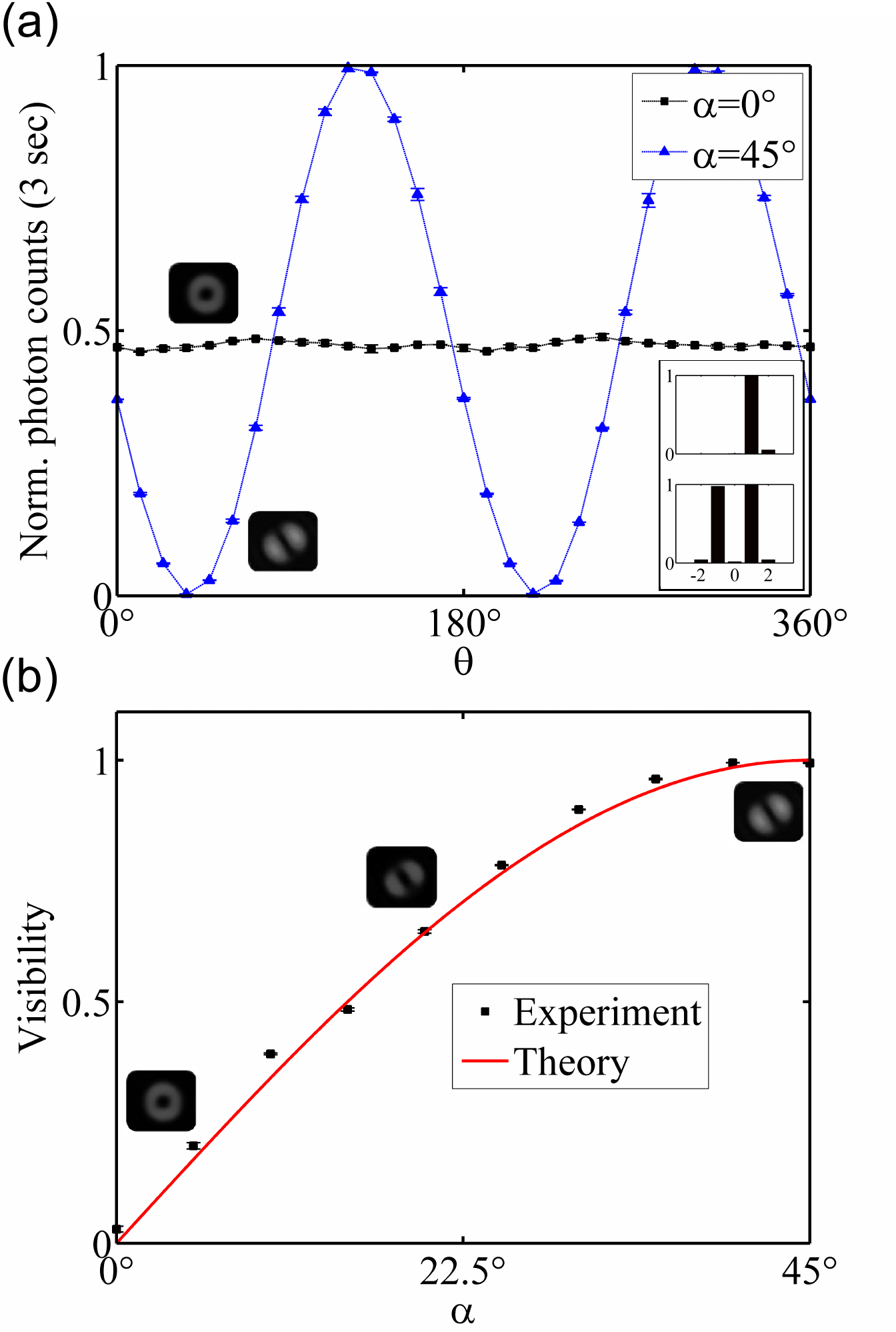}
	\caption{Experimental results for the free-space erasure showing that (a) the ``which-OAM'' information is distinguishable for $\alpha=0^\circ$ with visibility $V=0.02\pm0.06$ (squares) or indistinguishable for $\alpha=45^\circ$ (triangles) with $V=0.99\pm0.01$. The inset on the bottom right corner is the spectral decomposition of the OAM when $\alpha=0^\circ$ (top) and  $\alpha=45^\circ$ (bottom) in the range  $\ell=[-3,3]$ showing minimal mode cross-talk. In (b) the intermediate cases are investigated by varying the polarization projections from $\alpha=0^\circ$ to $45^\circ$ were the experimental visibility  increases as the OAM information is depleted.}	
	\label{fig:fig3}
\end{figure}

The detection of the hybrid mode of Eq.~(\ref{eq:hybrid1}) was carried out by projecting first the polarization DoF with a set of QWP, HWP and polarizer, followed by binary phase masks encoded on a SLM (Holoeye PLUTO) to project onto the particular spatial mode of the photon. This was done for $\alpha = [0^\circ,45^\circ]$, while scanning holograms through $\theta = [0^\circ,360^\circ]$. The projected photons were collected using a single-mode fiber and detected with avalanche photo-diodes (Perkin-Elmer).

The key part of the experiment consists in coupling the vector mode generated after the $q$-plate within the 30-mm-long step-index fiber core, and give an intuitive result based in the quantum eraser experiment, showing how much the hybrid state is affected by the disturbance introduced by the fiber. An objective (20$\times$) was used to improve the coupling of the probing mode into the fiber. Note that the following experimental results are a \emph{proof-of-concept}, that is why we use a short optical fiber channel for testing the technique.

\section{Results}

In the work presented here we explore the analogy between the path and the OAM degrees of freedom in a typical \emph{which-way} information experiment. In this particular case, the path information is encoded into an OAM mode, and thanks to its correlation with the polarization generated in the $q$-plate, the \emph{which-way} information can be erased by projecting the hybrid state in the diagonal polarization from Eq.~(\ref{eq:alpha}). Figure~\ref{fig:fig3} show the results for the best case scenario, when the vector mode is transmitted in free-space to give a reference curve, used later to probe the step-index fiber.

\begin{figure}[t]
	\centering \vspace{0.12cm}
	\includegraphics[width=8cm]{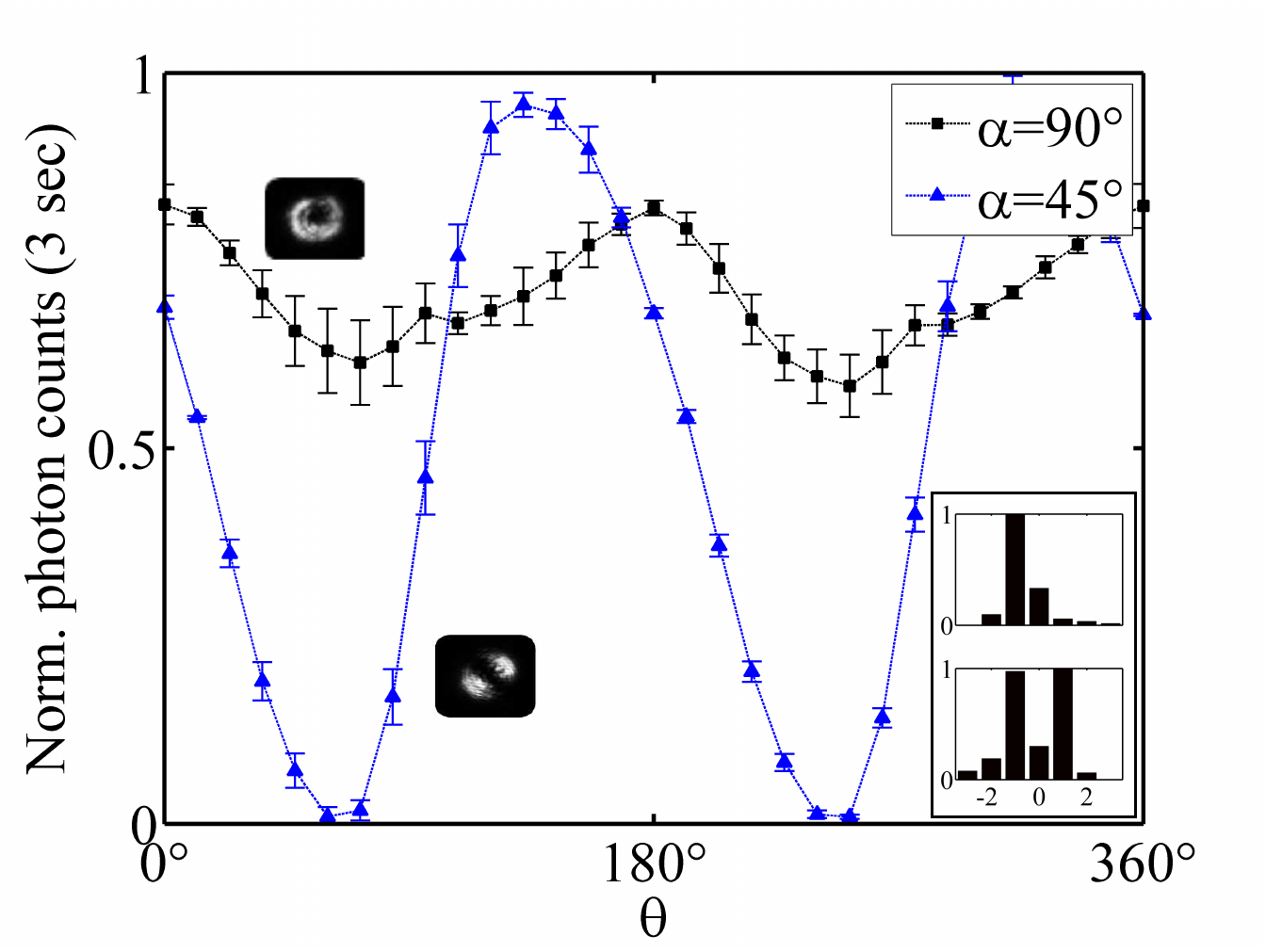}
	\caption{Experimental erasure results using the step-index fiber channel. The extreme case for $\alpha=90^\circ$ (squares) with a visibility $V=0.17\pm0.02$ and for $\alpha=45^\circ$ (triangles) with a visibility of $V=0.98\pm0.04$. The inset on the bottom right corner is the spectral decomposition of the OAM when $\alpha=90^\circ$ (top) and  $\alpha=45^\circ$ (bottom) in the range  $\ell=[-3,3]$.}	
	\label{fig:FiberResult}
\end{figure}

As can be seen in Fig.~\ref{fig:fig3}(a), the visibility of the spatial fringes that appear after projecting with the spatial superposition described in Eq.~(\ref{eq:theta}), can be maximized or minimized depending on the polarization projection $\alpha$ value from Eq.~(\ref{eq:alpha}). At $\alpha=45^\circ$ the projection value has maximum visibility, corresponding to the \emph{which-way} information maximally erased. At $\alpha=0^\circ$ the visibility is minimized due to the spatial superposition projection of a single OAM mode. That is to say that low visibility for the $\alpha=0^\circ$ case, corresponds to obtaining low spatial cross-talk between OAM modes within the given communication channel. Furthermore, the mode purity is confirmed by the spectral decomposition (as can be seen in the inset of Fig.~\ref{fig:fig3}(a)) where minimal mode cross-talk is observed. In addition, the inset confirms that the polarization projections lead to distinct OAM photon states when the OAM (path) is marked ($\alpha=0^\circ$) and a superposition when the OAM is erased ($\alpha=45^\circ$). Figure~\ref{fig:fig3}(b) shows the complete range of visibilities from the spatial fringes, with respect to the polarization projection $\alpha$ values. The maximum visibility obtained in the free-space configuration was $V=0.99\pm0.01$, and $V=0.02\pm0.06$ the minimum.

In Fig.~\ref{fig:FiberResult} we see from the spatial fringe curves that the fiber affects the hybrid transmitted state. From the visibility of the $\alpha=45^\circ$ case we can deduce the quality of erasing the \emph{which-way} information after the photon has traveled through the disturbing quantum channel. The maximum visibility obtained was $V=0.98\pm0.04$, which contrasts well with the reference free-space results, meaning that the cross-talk affects the spatial modes in a fairly symmetric manner. Thus, the spatial superposition is maintained as far as the good coupling into the step-index fiber core is preserved. On the contrary, when analyzing the results of $\alpha=90^\circ$ in Fig.~\ref{fig:FiberResult}, the minimum visibility curve ($V=0.17\pm0.02$) is poorer than the reference free-space curve from Fig.~\ref{fig:fig3}(b), implying that the spatial state projected after the step-index fiber is no longer a single OAM mode due to the cross-talk within the core, increasing the spatial superposition between OAM modes and increasing also the spatial fringes visibility correspondingly. Note that using $\alpha=90^\circ$ instead of $\alpha=0^\circ$, only changes the polarization projection but the cross-talk being analyzed remains the same. From Fig.~\ref{fig:FiberResult} we note that the mean number of photons detected in the case of $\alpha=90^\circ$ is not half of the normalized mean number of photons detected in the case of $\alpha=45^\circ$. This is because of alignment complications due to the laser instability and step-index miss-coupling after a few minutes.  Furthermore, the shape of the spatial fringes is perturbed due to the changes in the polarization and spatial DoF within the fiber. This is confirmed by the spectral decomposition (see Fig.~\ref{fig:FiberResult} inset) where modal cross-talk between OAM modes is observed further deteriorating the quality of the spatial modes. These visibilities and curves provide practical limits on what may be achieved in a quantum communication link down a fiber, and serve as a fast "pretest" to a full quantum key distribution or quantum state transfer down fiber.

The complete analogy between path information and spatial fringe visibility is essential to our quantum eraser scheme concept. By defining the two distinct paths using the OAM DoF, we have shown that through polarization-OAM hybrid state, it is possible to describe a simple projection scheme measurement capable of showing the total amount of disturbance present in a particular communication channel.

We point out that any DoF could be studied using this simple projective measurement, such as momentum, time-bin and even the frequency DoF. The polarization-OAM hybrid state is an attractive choice due to the possibility to explore the impact of spatial dimensionality in optical fiber communication links. That is, both DoF can be compensated before or after a particular disturbance in a communication channel by a unitary transformation, needing only a simple and fast detection scheme to implement a feedback measurement.  Previous works have shown that the non-separability of vector modes \cite{mclaren2015measuring} can be advantageous in measuring techniques for quantum entanglement to quantitatively  determine their mode quality in free-space \cite{ndagano2016beam} and fiber \cite{ndagano2015fiber}. Here we have performed a new channel measurement in fiber.

In conclusion, we have shown the complete analogy in the \emph{which-way} information concept, when using a path encoding approach or using instead a spatial DoF such as the OAM of a single photon. We have also derived a simple and fast projective detection scheme to measure the total impact of the environment on a polarization-OAM hybrid state, independent of the quantum channel used. We have focused our efforts in studying a step-index optical fiber channel due to the increasing relevance in high-order spatial encoding of information for ultra-fast fiber communications.

\section{Acknowledgments}
The authors express their gratitude to Lorenzo Marrucci for providing the $q$-plates. I.N. acknowledges financial support from the Department of Science and Technology (South Africa). C.R.G. and A.V. acknowledge support from the Claude Leon Foundation.

\bibliographystyle{apsrev4-1}

\end{document}